\documentclass[12pt,prd,preprint,tightenlines,superscriptaddress,nofootinbib]{revtex4-1}
\usepackage{graphicx}
\usepackage{amsmath,amssymb,mathrsfs}
\usepackage{bbm}
\usepackage{color}
\usepackage{ulem}
\usepackage{dsfont}
\usepackage{cancel}

\newcommand{\PRE}[1]{{#1}} 

\newcommand{\ben}{\begin{enumerate}}
\newcommand{\een}{\end{enumerate}}
\newcommand{\bit}{\begin{itemize}}
\newcommand{\eit}{\end{itemize}}

\newcommand{\beqa}{\begin{eqnarray}}
\newcommand{\eeqa}{\end{eqnarray}}
\newcommand{\beq}{\begin{equation}}
\newcommand{\eeq}{\end{equation}}
\newcommand{\bay}{\begin{array}}
\newcommand{\eay}{\end{array}}

\def\ifmath#1{\relax\ifmmode #1\else $#1$\fi}
\def\T{{\mathsf T}}
\def\gsim{\ \rlap{\raise 3pt \hbox{$>$}}{\lower 3pt \hbox{$\sim$}}\ }
\def\lsim{\ \rlap{\raise 3pt \hbox{$<$}}{\lower 3pt \hbox{$\sim$}}\ }

\def\Eq#1{Eq.~(\ref{#1})}
\def\eq#1{eq.~(\ref{#1})}
\def\eqs#1#2{eqs.~(\ref{#1}) and (\ref{#2})}

\def\ls#1{\ifmath{_{\lower1.5pt\hbox{$\scriptstyle #1$}}}}
\def\lsup#1{^{\lower 6pt\hbox{$\scriptstyle#1$}}}

\def\ord{{\cal O}}
\def\lt{\left}
\def\rt{\right}
\def\half{\tfrac{1}{2}}

\def\Im{{\rm Im}}
\def\Re{{\rm Re}}

\def\vev#1{{\langle #1\rangle}}
\def\Tr{{{\rm Tr}}}

\def\eg{e.g.}
\def\ie{i.e.}
\def\lag{{\cal L}}

\def\T{{\mathsf T}}

\def\bracket#1#2 {\mathinner{\langle{#1}|{#2}\rangle}}

\def\id{\mathds{1}}

\def\phm{\phantom{-}}

\begin{document}
\preprint{SCIPP 12/01, UCI-TR-2012-01}
\preprint{revised: October, 2012}
\vspace*{1cm}

\title{Group-theoretic Condition for Spontaneous CP Violation\PRE{\vspace*{.2in}}}

\author{Howard E. Haber}
\affiliation{Santa Cruz Institute for Particle Physics,
University of California, Santa Cruz, CA 95064, USA
\PRE{\vspace*{.2in}}
}

\author{Ze'ev Surujon\PRE{\vspace*{.2in}}}
\affiliation{Department of Physics and Astronomy, University of
California, Irvine, CA 92697, USA
\PRE{\vspace*{.2in}}
}

\begin{abstract}
\PRE{\vspace*{.3in}}
We formulate the necessary conditions for a scalar potential to exhibit spontaneous CP violation.
Associated with each complex scalar field is a U(1) symmetry that may be explicitly
broken by terms in the scalar potential (called spurions).
In order for CP-odd phases in the vacuum to be physical,
these phases must be related to spontaneously broken U(1) generators
that are also explicitly broken by a sufficient number of inequivalent spurions.
In the case where the vacuum is characterized by a single complex phase, our result implies that the phase must be
associated with a U(1) generator that is broken explicitly by at least two inequivalent spurions.
%
%
%
A suitable generalization of this result to the case of multiple complex phases has also been obtained.
These conditions may be used both to distinguish models capable
of spontaneous CP violation, and as a model building technique for
obtaining spontaneously CP-violating deformations of CP conserving models.
As an example, we analyze the generic two Higgs doublet model,
where we also carry out a complete spurion analysis.
We also comment on other models with spontaneous CP violation,
including the chiral Lagrangian, a minimal version of Nelson-Barr model,
and little Higgs models with spontaneous CP violation.
\end{abstract}

\maketitle

\section{Introduction}

In the Standard Model (SM), CP invariance is broken explicitly by
the Cabibbo-Kobayashi-Maskawa (CKM) phase.
Models beyond the SM often introduce additional CP-odd phases.
For example, these new sources of CP violation are needed to explain the
baryon asymmetry of the universe~\cite{bau}.
However, the observation of CP-violating phenomena does not necessarily imply
that the \textit{fundamental} source of CP non-invariance is due to the explicit breaking of CP.
In particular, CP-violating phenomena may be a consequence of
spontaneous CP violation, where the
Lagrangian of the theory respects the CP symmetry but the vacuum is not invariant under
CP.  Such a case arises when the vacuum expectation value (VEV) of a scalar field operator
exhibits physical CP-odd phases (which cannot be removed from the theory by field re-definitions).
Models in which all CP-odd phases, including the CKM phase, are due to
spontaneous CP violation, have the potential of solving the strong CP
problem, as exhibited by the Nelson-Barr models~\cite{NB}.

Explicit CP violation may be established by proving the non-existence of a
\textit{real basis}, \ie, a basis in field space where all couplings are real.
A basis-independent approach is one that identifies basis invariant quantities that are CP odd.
Perhaps the best-known example is the Jarlskog invariant~\cite{jarlskog} of the SM.
In contrast, the case of spontaneous CP violation is more complicated.
In its simplest form, spontaneous CP violation (SCPV) occurs
if and only if a class of real bases exists (which implies that the Lagrangian respects the CP symmetry),
but no real basis exists in which all the VEVs
are simultaneously real-valued.
In this case, the CP-odd invariants depend on the Lagrangian
couplings both explicitly and implicitly via the VEVs~\cite{DH,GH}.
Therefore, they are generally complicated functions of the model parameters.
Moreover, it is difficult to systematize the construction of these
invariants in a model-independent way.

It is the purpose of this paper to provide a model-independent formulation
of the necessary conditions for spontaneous CP violation.
These conditions derive from the fact that any phase in the VEV must be
related to a spontaneously broken U(1) generator.
In order for this phase to be physical, it is clear that the associated U(1)
symmetry should be broken explicitly.  The coefficients of the corresponding U(1) breaking terms that appear in the Lagrangian will
henceforth be called \textit{spurions}, since the explicit U(1) symmetry breaking
can be formally restored
by assigning appropriate transformation laws to these coefficients (in particular, spurions by definition carry a nonzero U(1) charge).
However, in order to guarantee that the phase in the VEV is physical,
there must be a sufficient number of inequivalent spurions relative
to the number of broken U(1) generators.  For example, a single complex
VEV can give rise to SCPV only if the associated U(1) is broken explicitly
by \textit{at least} two spurions whose U(1) charges differ in magnitude.
In theories of multiple complex scalars, there is a different U(1) associated
with each complex field.  Each spurion is characterized by a
\textit{charge vector}, whose components are the corresponding U(1) charges.
SCPV can arise only if the number of spurions $N_s$ is larger than the maximal
number of linearly independent charge vectors, denoted by $r$.  Moreover,
the number of potential CP-violating phases is determined to be equal to $r-r'$, where
$r'$ is the number of charge vectors that are linearly independent of the remaining
$N_s-1$ charge vectors.  A geometrical interpretation of this result is provided in Appendix A.

The practical implication of this formulation is three-fold.
First, it provides a simple way to find out whether a potential CP-odd phase in
the VEV is physical generically as a function of the model parameters.
Second, it provides a clearer way to understand why certain
regions in the parameter space never exhibit spontaneous CP violation
while others may do so.
Finally, given a CP-conserving model, our condition can be used to find a deformation
of that model which is spontaneously CP-violating in a generic region of its parameter space.

In this paper, we first compare and contrast explicit and spontaneous CP violation in Section II.
In Section III, we discuss in detail the necessary conditions for spontaneous CP violation.
We illustrate these conditions in Section IV by applying our results to the two
Higgs doublet model (2HDM)~\cite{tdlee,Branco:1985aq,Lavoura:1994fv,Botella:1994cs,DH,GH}.
In this analysis, the relevant explicitly-broken U(1) symmetry is the Peccei-Quinn symmetry~\cite{Fayet:1974pd,Peccei:1977hh}.
In Section V, we exhibit our conditions in other models of spontaneous CP-breaking,
by considering the chiral Lagrangian~\cite{dashen},
the minimal Nelson-Barr model~\cite{NB-minimal},
and the spontaneously CP violating Littlest Higgs~\cite{LLHCPV}.
Applying our formulation provides new insight to the question
of spontaneous CP violation in these models.
Our conclusions and future directions are given in Section VI.
In Appendix~B, we exhibit the full power of the spurion analysis for the 2HDM,
in which we examine spurions with respect the full SU(2) Higgs flavor group.  We reproduce results previously
obtained by Ivanov~\cite{Ivanov:2005hg}
and show how this formalism can be used for constructing
basis-independent invariants.
We also provide a more transparent understanding of the basis-independent condition for the existence
of the U(1) Peccei-Quinn symmetry in the 2HDM.

\section{Explicit and Spontaneous CP Violation}

The question of whether CP is violated explicitly or spontaneously deserves
some care due to the basis-dependence associated with the definition of CP.  For simplicity, we focus in this section
on scalar field theories, with scalar fields $\phi_i(\boldsymbol{\vec x},t)$, for $i=1,2,\ldots n$.
Consider the following generalized CP-transformation (GCP)~\cite{Ecker:1981wv,Ecker:1983hz,Neufeld:1987wa,Branco:1999fs,GCP,fhmns},
\beq \label{gcp}
\phi_i(\boldsymbol{\vec x},t)\longrightarrow X_{ij}\phi_j^\ast(\boldsymbol{-\vec x},t)\,.
\eeq
where $X$ is an $n\times n$ unitary matrix.  Such a transformation is automatically a symmetry of the
free scalar field theory action.  The form of this generalized CP transformation is basis-dependent.  Namely,
one can redefine the scalar fields such that $\phi^\prime_i(x)=U_{ij}\phi_j(x)$, where $U$ is an arbitrary
$n\times n$ unitary matrix.  The GCP-transformation in terms of the primed fields is
of the form given by \eq{gcp}, where $X$ is replaced by
\beq \label{uxu}
X^\prime =UXU^{\T}\,.
\eeq
The interacting scalar field theory is GCP-invariant if the action is invariant under \eq{gcp} for some choice
of $X$.  Three classes of GCP transformations exist: (i)~$XX^\ast=\mathds{1}$; (ii)~$XX^\ast= -\mathds{1}$;
and (iii)~$XX^\ast\neq \pm\mathds{1}$ (denoted in \cite{GCP,fhmns} as CP1, CP2 and CP3, respectively), where $\mathds{1}$ is
the $n\times n$ identity matrix.  However,
any CP2 or CP3 scalar field theory also respects CP1 (henceforth denoted as CP).  Hence, in what follows we focus on the case where $XX^\ast=\mathds{1}$,
which implies that $X$ is a symmetric unitary matrix.  We now employ the well known result that
any symmetric unitary matrix $X$ can be written as the product
of a unitary matrix and its transpose
(see e.g. Appendix D.3 of \cite{Dreiner:2008tw} for a proof of this result).
That is, one can always find a unitary matrix $U$ such that
$X=U^\dagger U^*$.  Using \eq{uxu}, it then follows that:
\beq \label{res1}
X^\prime\equiv UXU^{\T}=UU^\dagger (UU^\dagger)^*=\mathds{1}\,.
\eeq
That is, for any CP-invariant scalar field theory, there is always a basis choice for which
$X^\prime=\mathds{1}$, in which case the CP transformation reduces to complex conjugation and inversion of the space coordinate.

 {\bf 1. Explicit CP Violation (XCPV)}:
If a basis transformation $U$ can be found such that the scalar field theory action is invariant under \eq{gcp} with $X=\mathds{1}$,
then there exists a \textit{real basis}, \ie, a basis where all the couplings are real
and the model is explicitly CP conserving.  Conversely, if \eq{gcp} is \textit{not}
a symmetry of the scalar field theory action for any choice of the unitary matrix $X$, then
no real basis exists and the scalar field theory explicitly violates CP.

If the scalar field theory model is explicitly CP conserving, then a \textit{real basis} exists, with corresponding GCP transformation $X=\mathds{1}$.
Consider the set of basis transformations denoted by \{$U_r$\} that maintain the real basis.  This set necessarily includes
all real orthogonal $n\times n$ matrices.  Applying \eq{uxu}, we see that $X=\mathds{1}$ in a real basis related to the original
one by a real orthogonal basis change.  Depending on the form of the interacting scalar Lagrangian, the set \{$U_r$\} may also include
a subset of the unitary $n\times n$ matrices, denoted by \{$U_s$\}, that are not
real (and hence are not orthogonal).  In this case, the corresponding $X\neq\mathds{1}$.
Consider the ground state of a scalar field theory determined by a set of VEVs, $\langle\phi_i\rangle\equiv v_i$.
If the vacuum is GCP-invariant, then $v_i=X_{ij}v_j^\ast$.

{\bf 2. Spontaneous CP Violation (SCPV)}:
Given an explicitly CP-conserving scalar field theory, the vacuum is CP-invariant if and only if a real basis exists in which all the scalar
field VEVs are real (cf.~Theorem~3 in Appendix F of \cite{GH}).  Suppose that a real basis is chosen such that $X=\mathds{1}$ and the
scalar field VEVs are not all real.  It still may be possible to find a set of the basis transformations \{$U_s$\} that preserve the real basis such that
all the scalar field VEVs are real.  In this case, the scalar field theory and the vacuum are CP-conserving.
If the set $\{U_s\}$ is empty, then the model is said to exhibit spontaneous CP violation (SCPV).

Note that if the model is explicitly CP violating (\ie, there is no real basis:
the set $\{U_r\}$ is empty), then
the question of spontaneous CP violation is no longer meaningful, since there is no well-defined CP transformation law
that one can apply to the vacuum.

\section{Necessary Conditions for Spontaneous CP Violation}
\label{sec:conditions}

Given a real basis, spontaneous CP Violation is triggered by physical phases in the
VEVs.  We shall now examine what this implies for global symmetries and their
breaking.

\subsection{A Single Complex Scalar}
\label{sec:single}
Consider a complex scalar field degree of freedom, $\phi$, and the associated
field redefinition $\phi\to e^{i\alpha}\phi$.
This set of possible field redefinitions is a U$(1)$ subgroup of the maximal
global symmetry group O(2) of the kinetic energy terms.
Define the generator $X$ of this field redefinition, such that $\phi$ is
charged under U$(1)_X$ while the other degrees of freedom are neutral.
For certain potentials, the field $\phi$ acquires a VEV, $\vev{\phi}=ve^{i\theta}$,
breaking U$(1)_X$ spontaneously.
In this case, it is useful to parameterize the field in angular variables,
\beq
   \phi(x)=\rho(x) e^{iG(x)/v},
\eeq
where $G$ is a periodic field, $G\sim G+2\pi v$.
As long as U$(1)_X$ is not broken explicitly, $G$ is an exact Goldstone boson.
It shifts under U$(1)_X$ according to $G\to G+v\alpha$.
This induces a shift in the phase of the VEV, $\theta\to\theta+\alpha$, which defines
a circle of equivalent vacua.
Any phase $\theta_0$ is then unphysical, since it is equivalent to $\theta=0$ by a
U$(1)_X$ transformation which is an exact symmetry.

There are two possible ways to remove the Goldstone mode $G$
from the spectrum.
First, one may gauge U$(1)_X$, so that $G$ becomes the longitudinal
component of the associated gauge boson.
Note that in this case, the phase is still unphysical: it can be removed by a gauge
transformation.
A second possibility is to introduce explicit breaking of U$(1)_X$, such that
$G$ becomes a  massive pseudo-Goldstone boson.

An explicit breaking of U$(1)_X$ introduces a potential for the otherwise flat
Goldstone direction in field space.
Then one may ask whether $G$ acquires a VEV with a non-zero physical phase.
As a first attempt, suppose that U$(1)_X$ is broken by a single term in the
potential,
\beq
   V_X=\half b\phi^2+{\rm h.c.},
\eeq
where $b$ is real valued.
Apart from this term, the Lagrangian depends only on $\partial_\mu G$.
The new term introduces the only non-derivative dependence on $G$,
\beq \label{VX}
   V_X=b\rho^2\cos\frac{2G}{v},
\eeq
and is minimized at $\theta=\vev{G}/v=\pi/2$.
However, the phase can be removed by the field redefinition
$G\to G-v\pi/2$, which is equivalent to $\phi\to -i\phi$.
This transformation induces a sign flip, $b\to-b$,
such that in the new basis,
the minimum is at $\theta=0$ and there is no CP violation.

Had we introduced a different (higher) power of the field, \eg, $g\phi^4$, there would still be a
field redefinition (in this case: $\phi\to e^{-i\pi/4}\phi$) which removes the phase
from the VEV.  Note that while this transformation is not a symmetry,
it leaves the Lagrangian parameters real, merely changing the sign of $g$,
while removing the phase from the VEVs.
This is true for any single monomial $g_k\phi^k$.
The reason for this is that such a term always gives rise to a pure cosine potential,
$V_k(\theta)=2g_kv^k\cos(k\theta)$.
Since this potential has the property $V_k(\theta+\pi/k)=-V_k(\theta)$,
one can always choose a basis where
the minimum is at the origin, which implies that the vacuum conserves CP.

If we introduce two terms with different powers of $\phi$,
the resulting potential for $\theta$ becomes
a more general function, whose minimum cannot generically be shifted to the
origin without introducing a phase difference among the couplings.
Here the word ``generically'' should be interpreted as: ``in an $\ord(1)$ fraction of the
parameter space''.
As an example, consider
\beq \label{vx}
   V_X=b\phi^2+g\phi^4+{\rm h.c.},
\eeq
where $b$ and $g$ are real.
The new terms induce a potential for the otherwise flat $\theta$, which is given by
\beq \label{pot}
    V_X=bv^2\cos(2\theta)+gv^4\cos(4\theta).
\eeq
For parameters in the range $|b|<4gv^2$, this potential is minimized at
\beq \label{phase}
   \cos(2\theta_{\rm min})=-b/(4gv^2),
\eeq
generically resulting in spontaneous CP violation.

Although $V_X$ given in \eq{vx} provides an explicit violation of the U(1)$_X$ global
symmetry, we can formally make $V_X$ neutral under U(1)$_X$ by assigning two different U(1)$_X$
charges to the coefficients $b$ and $g$.  Indeed, if $b\to e^{-2i\alpha}b$ and
$g\to e^{-4i\alpha}g$, then $V_X$ is formally invariant under the U(1)$_X$
transformation $\phi\to e^{i\alpha}\phi$.  One can interpret $b$ and $g$ as vacuum
expectation values of two new scalar fields $\Phi_b$ and $\Phi_g$ respectively, in which case the explicit breaking
of U(1)$_X$ is reinterpreted as the spontaneous breaking of U(1)$_X$ due to the nonzero VEVs for the fields
$\Phi_b$ and $\Phi_g$.  In the literature, the VEVs $b\equiv\langle\Phi_b\rangle$ and $g\equiv\langle\Phi_g\rangle$ are commonly called \textit{spurions}.
Thus, in the above example, the spontaneous breaking of CP is attributed to the breaking of the
U(1)$_X$ symmetry by two spurions whose U(1)$_X$ charges differ in magnitude.

Note that for any spurion with U(1)$_X$ charge $q$, there is a complex conjugated spurion with
U(1)$_X$ charge $-q$.  Hence, it is the \textit{magnitude} of the charge that is relevant for
determining whether SCPV is possible.
Thus, we arrive at the following necessary condition for SCPV in the case of a single complex
scalar field:
\vskip 0.1in

\textit{Spontaneous CP violation in a theory of single complex scalar field may occur only if
the related U(1) is broken by at least two spurions whose U(1) charges differ in magnitude.}\\
\vskip 0.1in

Note that the value of the CP-violating phase in \eq{phase}
does not vanish in the $b,g\to 0$ limit, as long as $b/(gv^2)\sim 1$.
This may seem strange at first sight, but it can be understood as follows.
Without the explicit breaking, the phase is not physical and can take on any value.
With explicit breaking present, no matter how small, the phase becomes physical and its
value is stabilized by the effective potential.  That is, the explicit breaking terms
break the degeneracy of the unperturbed problem (in which the energy is independent of
the phase~$\theta$).  This is typical of all degenerate perturbation theory problems
in quantum mechanics.  Indeed, one can see that for $b/v^2$, $g\ll 1$
with $b/(gv^2)\sim 1$, the depth of the $\theta$-dependent part of the potential, \eq{pot},
is of order $gv^4\sim bv^2$.  Thus, the CP-violating phase becomes meaningless in
any physical process whose characteristic energy (or mass) is larger than $g^{1/4}v$.

\subsection{Multiple Complex Scalars}
%
In a model with multiple complex scalar fields, the vacuum
may be characterized by more than one CP-violating phase.
Although the value of any specific phase is basis-dependent, the number of potential\footnote{To determine whether a potential phase is physical, one must minimize
the effective potential of the phases to check for nontrivial solutions.}
CP-violating phases is well-defined and basis-independent.

The analysis of Section III.A shows that in
a model with a single complex scalar field,
the spurions are labeled by their U(1) charge and
SCPV requires at least two spurions with U(1) charges of different magnitude.
If this latter condition is satisfied, then the vacuum is characterized by at
most one independent CP-odd phase.  In the case of
$N$ complex scalar fields, the maximal symmetry group of the kinetic energy terms is O$(2N)$,
whereas the number of independent physical phases cannot exceed $N$.
These phases can always be taken
to be the ``diagonal'' phases associated with the Cartan subgroup
U$(1)_1\times\cdots\times$U$(1)_{N}$,
where each U(1) rotates the phase of one complex degree of freedom.%
\footnote{
Here we assume that none of the $N$ generators are gauged.  If some
of them are, the relevant group would be smaller.}
If the scalar potential contains
$N_s$ \textit{inequivalent} spurions, 
then each spurion may be labeled by an $N$-dimensional \textit{charge vector}
whose $j$th component is the charge under the U$(1)_j$.
Two spurions will be considered to be ``equivalent'' if their 
charge vectors are equal up to a possible overall
minus sign.\footnote{As previously noted, the charge vector of a complex conjugated spurion
is equal to the negative of the charge vector of a spurion.  Thus, we consider a spurion
and its charge conjugate to be equivalent in the present analysis.}

We construct the $N_s\times N$ matrix whose rows are given by the charge vectors of the spurions.
The rank $r$ of this matrix
is equal to the dimension of the vector space spanned by the corresponding charge vectors. 
Since the rank of a matrix cannot exceed the number of columns or rows, it follows that 
$r \leq{\rm min}~\{N_s,N\}$.  The physical interpretation of the rank is easily discerned.
Namely, only $r$ independent U(1)'s are broken by the spurions, which leaves
$N-r$ unbroken U(1)'s.   Hence, one can define new U(1) generators that are linear combinations of the original
U(1) generators such that the first $r$ U(1) generators are explicitly broken and the
last $N-r$ U(1) generators are unbroken.  In particular, the last $N-r$
components of the charge vectors of the spurions with respect 
to the new set of U(1) generators are zero. 

Thus, \textit{without loss of generality}, one can simply consider truncated $r$-dimensional
charge vectors (where the last $N-r$ zeros are removed).  Indeed, there can be at most $r$ physical
CP-violating phases associated with the $N$ complex scalar degrees of freedom, since
$N-r$ phases can be removed by employing the unbroken U(1)'s.
We shall denote the truncated $r$-dimensional charge vectors by
\beq \label{chv}
   \boldsymbol{q}^{(i)}\equiv\lt(q_1^{(i)},q_2^{(i)},\ldots,q_{r}^{(i)}\rt),\quad
   i=1,\ldots,N_s.
\eeq
As above, we can assemble the truncated charge vectors 
into an $N_s\times r$ matrix whose $i$th row is given by $\boldsymbol{q}^{(i)}$, which
we denote by $Q$.  By construction, $r={\rm rk}~\!Q$ and $N_s\geq r$.

Consider first the case where $N_s=r$.  This means that 
the $N_s$ vectors $\boldsymbol{q}^{(i)}$ are linearly independent and therefore
$Q$ is an invertible $r\times r$ matrix.  
It is convenient to redefine the U$(1)^{r}$ generators $\lt\{X_1,\ldots,X_{r}\rt\}$
by $X'_i\equiv \sum_j C_{ij}X_j$, where $C=(Q^{\T})^{-1}$.
Relative to this new basis for the U(1) generators, the charge vectors are given by
\beq \label{qprime}
   \delta^i_j=\sum_{k=1}^r C_{jk}q^{(i)}_k,\quad
   i,j=1,\ldots,r.
\eeq

Consequently, we have reduced the problem to $r$ independent copies of one complex scalar field and associated spurion (and its complex conjugate).  In particular, if we denote $\vev{\phi_n}=v_n e^{i\theta_n}$,
then the multi-field generalization of \eq{VX} is given by
\beq \label{vxprime}
V_{X'_1,X'_2,\ldots, X'_r}=\sum_{i=1}^{r}V_i(v_n)\cos\theta_i^\prime\,,\qquad
\text{where\,\, $\theta_i^\prime\equiv\sum_{k=1}^r q_k^{(i)}\theta_k$}\,,
\eeq
where $V_i(v_n)$ is the contribution to the potential of the $i$th spurion (where the
complex fields $\phi_n$ are replaced by the $v_n$, respectively).  
Using the results of Section III.A, we conclude that
no physical phases exist in the vacuum and thus there is no SCPV.  

In the case of $N_s>r$, we first label the truncated $r$-dimensional charge vectors such that
$\{\boldsymbol{q}^{(1)},\boldsymbol{q}^{(2)},\ldots,\boldsymbol{q}^{(r)}\}$ are linearly independent.
Then, the charge vectors of the remaining spurions, $\boldsymbol{q}^{(i)}$ for $i=r+1,r+2,\ldots,N_s$, are linear combinations of the first $r$ charge vectors.  
This means that that if we only keep the (inequivalent) spurions labeled by $i=1,2,\ldots,r$, we would again conclude that
no physical phases exist in the vacuum.  Hence, if we include \textit{all} $N_s$ inequivalent spurions,  
we are left with at least one potential physical phase.  To determine whether SCPV actually occurs, one must minimize the effective potential as in the single complex field case to determine
the vacuum value of this phase. We conclude the following:

\vskip 0.1in
\noindent
\textit{SCPV may occur only if the number of inequivalent spurions is larger than the dimension of the vector
space spanned by the corresponding charge vectors.}  
\vskip 0.1in

In a model of multiple complex scalar fields with $N_s>r$,
the number of potential physical phases (henceforth denoted by $d$) is obtained as follows.  
In analogy with \eq{qprime}, we define new charge vectors with respect to the redefined U(1)
generators $\{X_1,\ldots,X_r\}$,
\beq \label{Csum}
\sum_{k=1}^r C_{jk} q_k^{(i)}=\begin{cases} \delta^i_j\,, &\text{for $i=1,2,\ldots,r$}\,,\\[6pt]
q^{\prime\,(i)}_j\,,&\text{for $i=r+1,r+2,\ldots,N_s$}\,,\end{cases}
\eeq
where $C=(\widetilde{Q}^{\T})^{-1}$ and $\widetilde{Q}$ is the $r\times r$ matrix whose rows are the first $r$ (linearly independent) charge vectors $\{\boldsymbol{q}^{(1)},\boldsymbol{q}^{(2)},\ldots,\boldsymbol{q}^{(r)}\}$.  With respect to the redefined U(1), we can assemble the new charge vectors into an
$N_s\times r$ matrix, 
\beq \label{Qprime}
Q^\prime =\begin{pmatrix} \delta^i_j \\ ---- \\ q_j^{\prime\,(k)}\end{pmatrix}\,,
\eeq
where $i=1,2,\ldots, r$ and $k=r+1,r+2,\ldots,N_s$ label the $N_s$ rows of the matrix and $j=1,2,\ldots,r$.

One can now write out the spurion contributions to the scalar potential.  Using \eq{Csum}, the generalization
of \eq{vxprime} is immediate,
\beq \label{VXprime}
V_{X'_1,X'_2,\ldots, X'_r}=\sum_{i=1}^{r}V_i(v_n)\cos\theta_i^\prime
+\sum_{k=r+1}^{N_s} V_i(v_n)\cos\left(\sum_{j=1}^r q^{\prime\,(k)}_j\theta^\prime_j\right)\,,
\eeq
where $\theta^\prime_j$ is defined in \eq{Csum}.  The phases $\theta^\prime_j$ that explicitly appear in the second term of \eq{VXprime} are potential CP-violating phases.  Generically we would expect
$r$ CP-violating phases when $N_s>r$.  However, if there are $r'$ columns of zeros below the dashed
line in \eq{Qprime}, i.e. for all $k=r+1,\ldots,N_s$,
\beq
q_j^{\prime\,(k)}=0\,\,\text{for $r'$ values of the index $j$,}
\eeq
then only $r-r'$ phases appear in the second term of \eq{VXprime}.  The $r'$ phases  
that are absent do not acquire nontrivial CP-violating expectation values since for these phases,
the analysis reduces to the first case of $N_s=r$ treated above.  

There is a simple basis-independent
interpretation of $r'$.  Namely, $r'$ is equal to the number of charge vectors that are linearly
independent of the remaining $N_s-1$ charge vectors.  Thus, we conclude the following:
\vskip 0.1in
\noindent
\textit{For a scalar potential with $N_s$ spurion terms that exhibits SCPV, the number of potential CP-odd phases is given by $d=r-r'$.   That is $d$ is equal to the difference of the dimension of the vector space spanned by the $N_s$ charge vectors and the number of charge vectors that are linearly independent of the remaining $N_s-1$
charge vectors.}
\vskip 0.1in
\noindent
Note that the above result automatically incorporates the case of $N_s=r$ treated above, where 
all the charge vectors are linearly independent, in which case $r'=r$ and $d=0$.  That is, there is
no SCPV when $N_s=r$ as expected.  A geometrical
interpretation of the result $d=r-r'$ is given in Appendix A.

As a simple example
(which corresponds to the chiral Lagrangian of Section V.A), consider the
charge vectors $\{(1,0)\,,\,(0,1)\,,\,(-1,-1)\}$.  In this example, $N_s=3$ and
$N=r=2$.  However, note that none of the charge vectors is linearly independent of
the other two charge vectors.  In each case, we can express a given charge vector
as a linear combination of the other two.  Hence, in this example, $r'=0$ and
we conclude that $d=r-r'=2$.  
Thus, in this example there are two potential CP-violating phases that characterize
the vacuum.

If at least one of the $r-r'$ remaining nontrivial phases 
differs from a multiple of $\pi$ at the
minimum of $V_{X'_1,X'_2,\ldots, X'_r}$ [cf.~\eq{VXprime}], then the
model exhibits SCPV.  Generically, such a solution will exist if the
scalar potential parameters satisfy certain conditions.  In particular,
there will be a continuous range of scalar potential parameters that yields a continuous
range of values for the CP-violating phase(s).
Although we have implicitly assumed that the coefficients of
each spurion contribution to the scalar potential are independent, our
analysis also applies to cases in which the coefficients of 
inequivalent spurions are related due to, e.g., a discrete 
symmetry of the scalar potential.  In some scenarios of this kind, SCPV occurs
\textit{independently} of the choice of the remaining free 
scalar potential parameters (after the discrete symmetry is imposed),
in which case the corresponding CP-violating phases may take on only nontrivial discrete values.
An example of such a phenomenon is the so-called \textit{geometrical}
CP-violation of \cite{gt}. 

\section{Example: SCPV in the Two Higgs Doublet Model}
%
The two Higgs doublet model (2HDM) provides a good theoretical laboratory
for applying the results of the previous section.
Some of the results in this section are known.  Nevertheless, we reproduce
them here in a very simple and clear fashion by using our the group-theoretic
approach established in sec.~\ref{sec:conditions}.

The 2HDM consists of two hypercharge-one, SU$(2)_L$ doublets $(\Phi_1,\Phi_2)$.
The SU$(2)_L\times$U$(1)_Y$ gauge-covariant kinetic energy terms possess
an  SU$(2)_L\times$U$(1)_Y\times$SU$(2)_F$ symmetry, where the SU$(2)_F$ corresponds
to a ``Higgs-flavor'' symmetry transformation, $\Phi_i\to U_i{}^j\Phi_j$
with $U\in$\,\,SU(2)$_F$.  The generic 2HDM potential,
\beqa
   V &=& m^2_{11}\Phi_1^\dagger\Phi_1+m^2_{22}\Phi_2^\dagger\Phi_2
   -\lt(m^2_{12}\Phi_1^\dagger\Phi_2+{\rm h.c.}\rt)\nonumber\\
   &+&\half\lambda_1\lt(\Phi_1^\dagger\Phi_1\rt)^2+\half\lambda_2\lt(\Phi_2^\dagger\Phi_2\rt)^2
   +\lambda_3\Phi_1^\dagger\Phi_1\Phi_2^\dagger\Phi_2+\lambda_4\Phi_1^\dagger\Phi_2\Phi_2^\dagger\Phi_1
   \nonumber\\
   &+&\lt[\half\lambda_5\lt(\Phi_1^\dagger\Phi_2\rt)^2+\lambda_6\Phi_1^\dagger\Phi_1\Phi_1^\dagger\Phi_2
   +\lambda_7\Phi_2^\dagger\Phi_2\Phi_1^\dagger\Phi_2+{\rm h.c.}\rt],
   \label{eq:lag}
\eeqa
breaks the SU$(2)_F$ Higgs flavor symmetry completely.

Since there are four complex degrees of freedom, there are four potentially
physical SCPV phases, related to the four diagonal generators
\beq \label{t3}
   \id_{ij}\id_{\alpha\beta},\quad
   \id_{ij}T^3_{\alpha\beta},\quad
   T^3_{ij}\id_{\alpha\beta},\quad
   T^3_{ij}T^3_{\alpha\beta},
\eeq
acting on the $\Phi_{i\alpha}$, where SU$(2)_{F(L)}$ indices are denoted by
Roman (Greek) indices.
The first two are the diagonal generators of the
SU$(2)_L\times$U$(1)_Y$ gauge symmetry,
and thus cannot give rise to SCPV, as discussed in section~\ref{sec:conditions}.
As for $T^3_{ij}\id_{\alpha\beta}$, which generates the Peccei-Quinn (PQ)
symmetry~\cite{Fayet:1974pd,Peccei:1977hh} ($\Phi_{1}\to e^{i\alpha}\Phi_{1}$ and $\Phi_{2}\to e^{-i\alpha}\Phi_{2}$),
it is not gauged and is generically broken by the scalar potential.
Therefore it can potentially trigger SCPV.
The last generator $T^3_{ij}T^3_{\alpha\beta}$ (``chiral PQ'')
cannot give rise to SCPV in those vacua that preserve electric charge.
In particular the two VEVs are aligned in the U(1)$_{\rm EM}$ preserving vacuum, in which case chiral PQ
becomes degenerate with PQ.
%


In order to find models with SCPV, we choose a basis in which all the parameters in \eq{eq:lag} are real.
In this basis, we must then explicitly break the U$(1)_{\rm PQ}$.
We now perform a U$(1)_{\rm PQ}$ spurion analysis.\footnote{A more general SU(2)$_F$ spurion analysis is also
quite useful for other 2HDM applications.  See Appendix~B for further details.}
The various parameters transform formally under U$(1)_{\rm PQ}$ as follows.
The parameters $m_{11}^2$, $m_{22}^2$, and $\lambda_{1,2,3,4}$ are neutral with respect to U(1)$_{\rm PQ}$, whereas
the other parameters possess PQ charges:
\beq
   m^2_{12}[2],\quad \lambda_5[4],\quad\lambda_6[2],\quad\lambda_7[2],
\eeq
where we have assigned the fields with $\Phi_1[1],\ \Phi_2[-1]$.

In light of the above charge assignment, SCPV can arise in a realistic setting
only if
\ben
\item $\lambda_5$ is turned on.
\item At least one of the couplings $m_{12}^2$, $\lambda_6$, or $\lambda_7$
is turned on.
\item The other 2HDM parameters are chosen such that the SU$(2)_L\times$U$(1)_Y$ gauge symmetry is
broken to U(1)$_{\rm EM}$.
\een
Consider the following simple example (the general case is treated in Appendix B of \cite{gunhab}):
\beqa
   &&m^2_{11},m^2_{22}<0,\quad m^2_{12}=0,\nonumber\\
   &&\lambda_{1,2}>0,\quad \lambda_{5,6}\neq 0,\quad
   \lambda_3=\lambda_4=\lambda_7=0,
\eeqa
where $|\lambda_{5,6}|\ll \lambda_{1,2}$.
In this case,
\beq
   \langle\Phi_i\rangle\simeq \begin{pmatrix} 0\\ \sqrt{m^2_{ii}/\lambda_i}\end{pmatrix},
\eeq
with small corrections of order
$\ord\lt(\lambda_{5,6}/\lambda_{1,2}\rt)$.
We see that U$(1)_{\rm PQ}$ is broken only by the terms
\beq
   V_{\cancel{\rm PQ}}=\half\lambda_5\left(\Phi_1^\dagger\Phi_2\right)^2
   +\lambda_6\Phi_1^\dagger\Phi_1\Phi_1^\dagger\Phi_2+{\rm h.c.}
\eeq
We parametrize the two expectation values as
\beq
   \Phi_1^0=v_1e^{i\theta} e^{i\varphi},\quad
   \Phi_2^0=v_2e^{i\theta}e^{-i\varphi}.
\eeq
The new terms induce a potential for the otherwise flat $\varphi$, which is given by
\beq
    \Delta V= \lambda_5v_1^2v_2^2\cos(4\varphi)
    +2\lambda_6v_1^3v_2\cos(2\varphi).
\eeq
For parameters in the range
$|\lambda_6|\tan\beta<2\lambda_5$, this potential is
minimized at
\beq
   \cos(2\varphi_{\rm min})=\frac{\lambda_6}{2\lambda_5}\tan\beta,
\eeq
where $\tan\beta\equiv v_1/v_2$, resulting in spontaneous CP violation.

\section{Other Models of Spontaneous CP Violation}

In this section, we briefly examine other models that exhibit SCPV,
in light of the necessary conditions developed in Section III.

\subsection{The Chiral Lagrangian}
\label{sec:dashen}
Dashen's model of spontaneous CP violation~\cite{dashen}
is based on the three-flavor chiral Lagrangian  (see e.g.~\cite{creutz} for a
modern review).
Recall that this theory is the low energy description of three-flavor QCD, and it
describes the spontaneous breaking of
\beq \label{breaking}
   {\rm SU}(3)_L\times{\rm SU}(3)_R\to {\rm SU}(3)_V,
\eeq
where the two SU(3) groups act on the left- and right-handed quarks $(u,d,s)$, respectively.
The vacuum transforms as $(3,\bar 3)$ under
${\rm SU}(3)_L\times{\rm SU}(3)_R$:
\beq \label{sigzero}
    \Sigma_0\to
    L(\varepsilon_L^a)\,\Sigma_0\, R^\dagger\!(\varepsilon_R^b)\,.
\eeq
In order to ensure that only the diagonal SU$(3)_V$ transformations ($L=R$)
leave the vacuum invariant as required by \eq{breaking}, it follows that $\Sigma_0=\id$.

Note that the condition $\Sigma_0=\id$ is basis dependent.  Indeed, one can simply redefine
all $(3,\bar{3})$ fields by applying an arbitrary ${\rm SU}(3)_L\times{\rm SU}(3)_R$ transformation.
As a result of such a field redefinition,
\beq \label{U}
\Sigma_0=U\,, \qquad U\in~{\rm SU(3)}\,.
\eeq
Relative to the new basis, the symmetry-breaking pattern is
${\rm SU}(3)_L\times{\rm SU}(3)_R\to {\rm SU}(3)_U$, where an SU(3)$_U$ transformation corresponds to
$R=U^\dagger LU$ in \eq{sigzero}.

As a consequence of the spontaneous breaking of chiral symmetry, there
are eight Goldstone modes $G^a=\{\pi^i,\,K^i,\,\eta\}$, which are
parameterized as
\beq \label{gold}
   G(x)\equiv G^a(x)T^a
   =\frac{1}{\sqrt{2}}\begin{pmatrix}
      \frac{1}{\sqrt{2}}\pi^0+\frac{1}{\sqrt{6}}\eta & \pi^+ & \phm K^+\\
      \pi^- & -\frac{1}{\sqrt{2}}\pi^0+\frac{1}{\sqrt{6}}\eta & \phm K^0\\
      K^- & \overline{K^0} & -\frac{2}{\sqrt{6}}\eta\end{pmatrix},
\eeq
where the $T^a$ are the SU(3) generators in the fundamental representation.
The chiral Lagrangian is expressed in terms of the $(3,\bar{3})$ field $\Sigma(x)$, which depends on the Goldstone fields via
\beq
\Sigma(G)=e^{iG(x)/f}\Sigma_0 e^{iG(x)/f}\,,
\eeq
where $\Sigma_0\equiv\langle\Sigma\rangle$.
In the case of $\Sigma_0=\mathds{1}$,
the Goldstone fields transform linearly under the vector SU(3)$_V$ and transform nonlinearly
and non-homogeneously under the spontaneously broken axial transformations,
for which $L=R^\dagger$.
The non-homogeneous term of the transformation law is a signal that the Goldstone
fields are massless and derivatively coupled,
as long as there are no explicit SU$(3)_L\times$SU$(3)_R$ breaking terms in the Lagrangian.
Of course, these conclusions do not depend on the choice of $\Sigma_0=\mathds{1}$, since all
vacua related by the matrix $U$ in \eq{U} are equivalent.

However, in order for the chiral Lagrangian to describe nature, the chiral symmetry
must be broken explicitly. Such explicit breaking is introduced both by
electromagnetic gauge interactions and by the quark masses.
The chiral Lagrangian takes the form
\beq \label{chlag}
   \lag=\tfrac{1}{4}f^2{\rm Tr}
   \lt(\mathcal{D}_\mu\Sigma^\dagger\mathcal{D}^\mu\Sigma\rt)
   +\half B_0 f^2{\rm Tr}\left(M\Sigma^\dagger +\Sigma M^\dagger\right),
\eeq
where $B_0$ is proportional to the quark-antiquark condensate (see, e.g., \cite{dsm}),
\beq
   M=\begin{pmatrix} m_u &\,\,\,0 & \,\,\, 0\\ 0&\,\,\, m_d & 0 \cr \,\,\, 0& \,\,\, 0 &\,\,\, m_s\end{pmatrix},
\eeq
and $\mathcal{D}_\mu$ is the gauge covariant derivative.
Once explicit chiral symmetry-breaking is introduced, all vacua related by the matrix $U$ in \eq{U} are
no longer equivalent.  In particular, vacua corresponding to different eigenvalues
of $U$ are now inequivalent (e.g.~they have different energy values).
For example, if the quark masses are all positive, then the potential energy due to the explicit
chiral symmetry breaking is minimized by assuming that $\Sigma_0=\mathds{1}$.
However, it is possible that some of the quark mass parameters are negative.\footnote{The physical quark masses are given by the absolute values of the quark mass parameters.  Nevertheless, the signs of the
quark masses can have physical relevance, as the present discussion makes clear.}
Without loss of generality, one can choose the vacuum value $\Sigma_0=U$ to be diagonal.
Since $U\in$~SU(3), the diagonal elements are pure phases whose product is equal to one.
That is,
\beq
   \Sigma_0=
   \begin{pmatrix} e^{i\theta_u} & \,\,\, 0&\,\,\, 0 \\
 0  &\,\,\,  e^{i\theta_d} &\,\,\, 0 \\
 0  & \,\,\, 0& \,\,\,  e^{-i(\theta_u+\theta_d)}\end{pmatrix}.
\eeq
Dashen's observation was that a region exists in the $(m_u,m_d,m_s)$
parameter space
where $\theta_u$ and $\theta_d$
are not minimized at the origin, thus inducing SCPV.
The potential for the phases is
\beq
   V=B_0 f^2
   \lt[m_u\cos\theta_u+m_d\cos\theta_d+m_s\cos\lt(\theta_u+\theta_d\rt)\rt].
\eeq
Provided that $m_u m_d<0$,\footnote{For $m_u m_d>0$, the extremum condition given by \eq{min} is a local maximum.}
the potential above is minimized when~\cite{creutz}
\beq \label{min}
   m_u\sin\theta_u=m_d\sin\theta_d=-m_s\sin\lt(\theta_u+\theta_d\rt)\,.
\eeq
%
%
\begin{figure}[t!]
   \centering
   \includegraphics[width=0.6\textwidth]{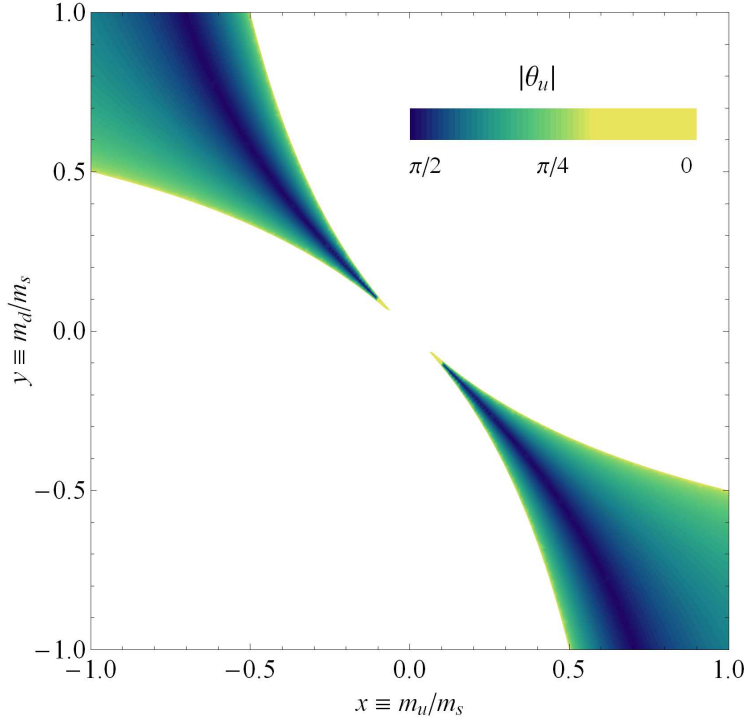}
   \caption{Regions of the parameter space of Dashen's Model
   parameter space of Dashen's model, where spontaneous CP violation
   occurs [cf.~\eq{ineq}].  A point in this parameter space corresponds to
   $(x,y)\equiv(m_u/m_s,m_d/m_s)$.
   The size of the phase $\theta_u$ is shown, with maximum values depicted
   in dark (blue), and minimum values in light (yellow).
   In these regions, $\theta_d$ also acquires a nonzero value, as explained
   in the text. The value of $\theta_d$ at the point $(x,y)$ is equal to the
   value of $\theta_u$ at the point $(y,x)$.
\label{fig:dashen}}
\end{figure}

It is convenient to introduce dimensionless mass ratios,
\beq
x\equiv m_u/m_s\,,\qquad y\equiv m_d/m_s\,.
\eeq
Assuming $xy<0$, we can use \eq{min} to obtain the vacuum values of $\theta_u$ and $\theta_d$,
\beq \label{p1}
   \cos\theta_u=
   \tfrac{1}{2}\lt(\frac{y}{x^2}-\frac{1}{y}-y\rt)\,,\qquad\quad
\cos\theta_d=
   \tfrac{1}{2}\lt(\frac{x}{y^2}-\frac{1}{x}-x\rt)\,,
\eeq
under the assumption that $-1\leq\cos\theta_{u,d}\leq 1$.  If this latter assumption is false, then
the minimum of the potential for the phases lies on the boundary where $|\cos\theta_{u,d}|=1$,
corresponding to a CP-conserving vacuum.  Thus, SCPV can arise if and only if $xy<0$ and
$-1<\cos\theta_{u,d}<1$.  Using \eq{p1}, these inequalities yield~\footnote{Note that if
we interchange $x$ and $y$ in \eq{ineq}, the results are identical to the original inequalities.}
\beq \label{ineq}
\frac{|x|}{1+|x|}< |y|< \frac{|x|}{1-|x|}\,,\qquad xy<0\,,
\eeq
in which case the vacuum is characterized
by two independent physical phases $\theta_u$ and $\theta_d$ given by \eq{p1}.
In Fig.~\ref{fig:dashen}, we show regions of the $x$--$y$ plane that
admit SCPV.   Indeed, this range is ruled out phenomenologically (using the light quark masses quoted in \cite{pdg}).

Although Dashen's model is no longer a viable model for CP-violation, we can use this model to
illustrate the results of Section~III.B in the case of more than one U(1) factor.
Prior to turning on the explicit breaking terms (namely the spurions $m_u$, $m_d$ and $m_s$),
there are two spontaneously-broken U(1) generators that can be identified
with the two diagonal SU(3) generators $T^3$ and $T^8$.  In fact, it is more convenient to define
linear combinations of these two generators,
\beq
T_u\equiv T^3+\sqrt{3}\,T^8=\begin{pmatrix} 1 & \,\,\,\phm 0 &  \,\,\,\phm 0 \\ 0 &  \,\,\,\phm 0 &  \,\,\,\phm 0 \\ 0 &  \,\,\,\phm 0 & \,\,\, -1\end{pmatrix}\,,
\qquad\quad
T_d\equiv -T^3+\sqrt{3}\,T^8=\begin{pmatrix} 0 & \,\,\, \phm 0 & \,\,\, \phm 0 \\ 0 &  \,\,\,\phm 1 & \, \,\,\phm 0 \\ 0 &  \,\,\,\phm 0 & \,\,\, -1\end{pmatrix}\,,
\eeq
which can be used to shift the values of $\theta_u$ and $\theta_d$, respectively.
Applying $T_u$ and $T_d$ to the vectors $(1,0,0)$, $(0,1,0)$ and $(0,0,1)$
yields the U(1)$_u$ and U(1)$_d$ charges of the three spurions, respectively.
The corresponding charge vectors are given by:
\beq
   m_u(1,0),\quad m_d(0,1),\quad m_s(-1,-1).
\eeq
The three charge vectors are linearly dependent and span a two-dimensional vector space.
In the notation of III, we have $N_s=3>{\rm rk}~\!Q=2$, in which case SCPV is possible.
Indeed the conditions for SCPV derived in Section III.B,
when applied to the above set of spurions, yields
potentially two independent physical CP-violating phases $\theta_u$, $\theta_d$ that characterize the vacuum.
%

Had we considered a chiral Lagrangian based on U$(3)_L\times$U$(3)_R$ instead of SU$(3)_L\times$SU$(3)_R$,
then $\Sigma_0={\rm diag}(e^{i\theta_u}\,,\,e^{i\theta_d}\,,\,e^{i\theta_s})$,
with no relation among the three phases.  Prior to turning on the explicit breaking terms,
there are now three spontaneously-broken U(1) generators that can be identified
with $T_u$, $T_d$ and $T^0$, where $T^0$ is the $3\times 3$ identity matrix which generates
an axial U(1)$_A$ transformation.  The corresponding charge vectors of the spurions,
\beq
   m_u(1,0,1),\quad m_d(0,1,1),\quad m_s(-1,-1,1)\,,
\eeq
are linearly independent, spanning the full three-dimensional vector space, so that $N_s\!=\!{\rm rk}~Q$.
Naively, it seems that none of the three phases is physical,  resulting in the absence of SCPV.
However, the axial U(1)$_A$ symmetry is anomalous, and can be modeled
by adding an explicit U(1)$_A$ breaking term to the chiral Lagrangian that is proportional to
$(\ln\det\Sigma)^2$~\cite{Rosenzweig:1979ay,DiVecchia:1980ve,Witten:1980sp}.
Consequently,  there is a fourth spurion so that $N_s=4>{\rm rk}~\!Q=3$, and we again conclude that SCPV is possible.
The corresponding fourth charge vector is $(0,0,1)$; hence
the analysis of Section III implies that there are three potential physical CP-violating
phases $\theta_u$, $\theta_d$ and~$\theta_s$ that characterize the vacuum.
Hence, including the axial U(1)$_A$ symmetry and its anomaly-induced explicit breaking
does not spoil the existence of a SCPV phase in the parameter space of the
chiral Lagrangian.  A more detailed study is presented in~\cite{Witten:1980sp}, where the effect of the
strong CP angle $\theta$ is also taken into account.

Finally, it is noteworthy that in the case of two light flavors, the effective potential of the SU$(2)_L\times$SU(2)$_R$ theory
depends only on $\cos\theta_u=\cos\theta_d$, whereas the corresponding
spurions are $m_u(1)$ and $m_d(-1)$.  Using the nomenclature of Section III, there is only one inequivalent spurion,
in which case the SU$(2)_L\times$SU(2)$_R$ chiral Lagrangian cannot give rise to SCPV.

\subsection{The Minimal Nelson-Barr Model}
Here we will consider the model by Bento, Branco, and Parada~\cite{NB-minimal}.
The model solves the strong CP problem by imposing CP as an exact symmetry, and breaking
it spontaneously, thereby producing unsuppressed CKM phase, along with a suppressed
strong CP phase.

The field content of the model is the SM plus one gauge singlet
complex scalar $S$, and one pair of vector-like down quarks $D_L,D_R$.
The new interactions of the Lagrangian are given by
\beq
   \delta\lag=-\mu\overline{D}_LD_R-(f_iS+f'_iS^*)\overline{D}_Ld_R^i+{\rm h.c.}
\eeq
Moreover, due to the presence of terms such as $S^2$, $S^4$, etc.,
there is a range of parameter space for which $\vev{S}=Ve^{i\alpha}$.
The phase $\alpha$ eventually feeds into the SM fermion mass matrices and provides
the sole source of CP violation.  Since the couplings $f_i$ and $f'_i$ are flavor-dependent,
this phase can become the CKM phase once both scalars acquire VEVs.
The radiatively induced strong CP-violating parameter $\bar\theta$
is small and therefore the strong CP problem is solved.

In terms of our group theoretical condition, the scalar sector has a single spontaneously
broken U(1) that is explicitly broken by more than one spurion (e.g., $S^2$ and $S^4$), such that the vacuum has
one physical nonzero phase.  This is similar to the toy model with one complex scalar field presented
in section~\ref{sec:conditions}.

\subsection{Little Higgs models}

The Little Higgs framework is a class of nonlinear sigma models
that produce the SM as their low energy limit.
By careful design, dubbed ``collective symmetry breaking''~\cite{LH},
the Higgs mass parameter does not receive quadratically divergent
corrections at one-loop.
These models can potentially solve the little hierarchy problem, since it allows for the Higgs mass
to be of $\ord$(100 GeV) even when
the UV cutoff is as high as $\ord$(10~TeV).

A popular little Higgs model is the Littlest Higgs model of~\cite{LLH},
in which SU$(5)$ is broken to SO$(5)$ by a two index symmetric SU$(5)$ tensor.
The Lagrangian is given by
\beq
   \lag=\frac{f^2}{8}{\rm Tr}\lt|{\cal D}_\mu\Sigma\rt|^2
   +\lambda_1 f\bar Q_i\Omega^i t_R+\lambda' f\bar t'_L t'_R+{\rm h.c.},
\eeq
where $f$ is the Goldstone decay constant, $Q_i$ and $t'_{L,R}$ are fermions,
and $\Omega^i$ is an SU$(5)$ breaking function of $\Sigma$ elements that
is chosen in accordance with the principle of collective symmetry breaking.

In a variant of this model~\cite{PPP}, there is an exact global U$(1)$
which is spontaneously broken~\cite{LLHCPV}.
This U$(1)$ is generated by $Y'={\rm diag}(1,1,-4,1,1)$.
As a result, there is an exact Goldstone mode $\eta$ associated with $Y'$.
In order to make the theory viable, the field $\eta$ must acquire mass,
requiring explicit breaking of U$(1)_{Y'}$.
A possible spurion that breaks this U$(1)$ would be
$s=(0,0,1,0,0)^T$, transforming (formally) under the fundamental representation
of SU$(5)$.
Its symmetry breaking pattern is SU$(5)\to$~SU$(4)$, which acts on the
$(3,3)$ minor.
The nine broken generators include $Y'$ and generators which are also
broken by the gauging.
In particular, any function of $\Sigma_{33}=s^\dagger\Sigma s$
would break $Y'$ while maintaining gauge invariance.
The term,
\beq
   \delta\lag=\varepsilon f^4\Sigma_{33}+{\rm h.c.},
   \label{eq:sigma33}
\eeq
is sufficient to generate mass for the Goldstone boson $\eta$ [cf.~\eq{chlag}].
However,  a physical CP-odd
phase can arise only in the presence of at least two different terms.
As a simple example, consider
\beq
   \delta\lag_{\rm SCPV} =
   \varepsilon f^4\lt(a\Sigma_{33}+b\Sigma_{33}^2\rt)+{\rm h.c.},
\eeq
where we take $\varepsilon,a,b$ to be real, with $a,b\sim\ord(1)$ and
$\varepsilon$ loop-suppressed.
This results in the following tree-level potential for $\eta$:
\beq
   V_\eta=2\varepsilon f^4\lt(a\cos\frac{2\eta}{\sqrt{5}f}
   +b\cos\frac{4\eta}{\sqrt{5}f}\rt).
\eeq
This potential is minimized for
\beq
   \vev{\eta}=\half\sqrt{5}f\arccos\lt(\frac{-a}{4b}\rt)
   \qquad {\rm if}\quad \lt|\frac{a}{4b}\rt|<1,
\eeq
which is of order one if we assume no hierarchy between $a$ and $b$.

Further discussion of CP violation in this class of models and related issues
can be found in~\cite{LLHCPV}.

\section{Conclusions and Future Directions}

We have formulated the necessary conditions for spontaneous CP violation
from a group-theoretic perspective, \ie, in terms of breaking patterns
of global U(1) symmetry generators.
This new framework allows for a more systematic study of
spontaneous CP violation model building.
We have used the fact that CP-violating phases in the vacuum are
related to operators that explicitly break the corresponding U(1) groups
and the corresponding spurions which are the coefficients of these operators.
Such phases are nontrivial and signal spontaneous CP-violation only
in cases where there are a sufficient number of inequivalent spurions relative
to the number of broken U(1) generators.

We assume that the scalar potential of the model is explicitly CP-conserving.
In the case of a single CP-violating phase that characterizes the vacuum, the phase is physical only if the associated
U(1) is broken explicitly by \textit{at least} two spurions whose U(1) charges differ
in magnitude.  We have generalized this result to the case of multiple phases and the associated
U(1) factors.  To each spurion, one can assign a charge vector whose components are
the U(1) charges. 
Two spurions
are called equivalent if their charge vectors are equal (up to a possible overall minus sign).
If there are $N_s$ inequivalent spurions whose charge vectors span an $r$-dimensional vector subspace,
then there is at least one potential physical CP-violating phase that characterizes the vacuum only if $N_s>r$.
The number of potential CP-odd phases is then determined to be equal to $r-r'$, where $r'$ is
the number of charge vectors that are linearly independent of the remaining $N_s-1$ charge vectors.
%
The actual value of the potential CP-violating phase is ultimately determined by minimizing an effective
potential.  If a minimum exists
such that at least one CP-violating phase is $\theta_{\rm CP}\neq 0,\pi$, then
the CP symmetry is spontaneously broken.

Using these results, we have analyzed the two Higgs doublet model, Dashen's model
for spontaneous CP violation in the chiral Lagrangian, a minimal Nelson-Barr
model, and the Littlest Higgs with spontaneous CP violation.
For the two-Higgs doublet model, we have also performed a comprehensive spurion analysis,
in which we employ the full SU(2) Higgs flavor group.  We reproduce results previously
obtained by Ivanov~\cite{Ivanov:2005hg}, and demonstrate how to use this formalism to
construct invariant relations that are independent of the choice of scalar field basis.

The applications presented in this paper focus on tree-level results.  It is of interest to consider
whether our framework allows for spontaneous CP-violation to be generated by radiative effects.
Consider the case of a single CP-violating phase that characterizes the vacuum.  For this to
be a robust result that holds over an $\mathcal{O}(1)$ fraction of the model parameter space, one
requires the two inequivalent spurions to be of comparable size.  If one of the spurions arises from
a tree-level operator and the other arises radiatively, then it appears that the latter requirement
cannot be satisfied (without violating perturbativity of the loop expansion).

Nevertheless, one can
imagine a number of scenarios in which spontaneous CP-violation is radiatively generated.  For example,
in a model with multiple complex scalars, it may be possible to radiatively generate two inequivalent
spurions at the loop level, which could result in an $\mathcal{O}(1)$ CP-violating phase.
Alternatively, the tree-level spurions might arise from a different sector of the theory (such
as the fermion sector), in which case one could balance that against a radiatively generated spurion
in the scalar sector.  However, the Georgi-Pais theorem~\cite{Georgi:1974au} 
limits the ways in which CP violation can be induced radiatively, without
introducing unnaturally light scalars.

Finally, we note that some of the the global U$(1)$ symmetries related to the CP-violating phases may be anomalous.
In this case, the anomaly is manifested by the presence of explicitly breaking terms in the Lagrangian.
If the terms generated by the anomaly satisfy the necessary conditions developed in this paper,
then one could imagine the possibility of spontaneous CP-violation whose presence is due to the anomaly.
It would be instructive to find explicit models that realize this possibility.
We leave these interesting possibilities for a future study.

\begin{acknowledgments}

We appreciate useful discussions with Gustavo Branco, Benjamin Grinstein, 
Richard Hill, Aneesh Manohar, Patipan Uttayarat and Ivo de Medeiros Varzielas.  
We are especially grateful to an anonymous referee
who encouraged us to provide a more transparent derivation of the number of CP-violating
phases established in Section III.B.  The final stages of this work was supported in part
by the National Science Foundation under Grant No. PHY-1055293 and the hospitality of the
Aspen Center for Physics.  H.E.H. is supported in part by U.S. Department of
Energy grant number DE-FG02-04ER41268.

\end{acknowledgments}

\appendix

\section{Geometrical interpretation of the number of CP-violating phases}

We can employ the following geometrical construction for establishing the number of potential
CP-violating phases in a SCPV scalar potential.  Using the notation of Section III.B, we first
select a linearly independent set of $r$ charge vectors and
use this set as a basis for the linear space of charge vectors. 
This basis can be used to construct hyperplanes spanned by a subset of the basis vectors.  For example,
each basis vector defines a one-dimensional line that lies parallel to the corresponding basis vector, each pair
of basis vectors spans a two-dimensional plane, etc.  Consider the set of
all such hyperplanes.  From this set, we can identify the unique hyperplane of minimal dimension $d$ 
that contains the span of the remaining $N_s-r$ charge vectors.   
(Note that $d$ must lie in the range
$1\leq d\leq r\leq N$.)  Each basis vector that lies in
the hyperplane of minimal dimension is associated with a physical phase.  For example, if the remaining charge vectors
are all parallel to a single basis vector then $d=1$, in which case there is one potential
physical phase.  We conclude the following:

\vskip 0.1in
\noindent
\textit{The number of potential CP-violating phases is equal to $d$, obtained by determining the
unique hyperplane of minimal dimension $d$, constructed from all possible subsets of the $r$ basis vectors,
in which the span of the remaining $N_s-r$ charge vectors resides.}
\vskip 0.1in

\noindent

The procedure presented above is inherently geometrical.  In particular,
the number $d$ does not depend on the initial choice of the $r$ linearly-independent basis vectors.
Hence, the number of potential CP-violating phases that characterizes the vacuum is a basis-independent concept.
Indeed, it is straightforward to show that this procedure yields the result obtained in Section III.B.  In particular,
it is convenient to employ the basis of U(1) generators that yields the matrix $Q'$ given in \eq{Qprime}. 
Let $r'$ be the number of columns of zeros that lie below the dashed line in \eq{Qprime}.  Focusing
on the remaining $r-r'$ columns, consider the row vectors whose 1 appears in one of these $r-r'$
columns.  The span of these row vectors is a hyperplane of dimension $d=r-r'$, which we identify 
as the number of potential CP-violating phases. 

As a simple example, we again consider the charge vectors 
$\{(1,0)\,,\,(0,1)\,,\,(-1,-1)\}$, where $N_s=3$ and $N=r=2$.  
For any $c\neq 0$, the vector $c(-1,-1)$ is neither
parallel to $(1,0)$ nor to $(0,1)$.  Indeed $c(-1,-1)$ lies in the two dimensional
plane spanned by $(1,0)$ and $(0,1)$, so that hyperplane of minimal dimension that 
contains $c(-1,-1)$ is a plane of dimension $d=2$.
Thus, in this case there are two potential CP-violating phases that characterize
the vacuum.

Note that the above analysis applies trivially to the case of a single complex scalar field, where $N=r=d=1$.  
In this case, there is one potential CP-violating phase if $N_s>r$, which yields $N_s>1$.  
This conclusion coincides with the analysis given in Section III.A for the case of one complex scalar field.

\section{The 2HDM spurion analysis and some applications}

\subsection{Full SU$(2)_F$ spurion analysis}

We begin by expressing all the parameters in the 2HDM scalar potential in terms of invariants and
spurions
of SU$(2)_{F}$.
This is accomplished by constructing gauge invariant terms
from
the fields $\Phi_{i\alpha},\ \overline\Phi^{i\alpha}\equiv (\Phi_{i\alpha})^\dagger$ and the invariants
$\epsilon_{ij},\ \epsilon_{\alpha\beta},
\ \delta_\alpha^\beta,\ \delta_i^j$~\cite{Velhinho:1994vh,
Ivanov:2005hg,Maniatis:2006jd,Nishi:2006tg, Ivanov:2006yq}.
The only gauge-invariant bilinear term is
$(M^2)^{\ i}_{j}\overline\Phi_{i\alpha}\Phi^{j\alpha}$, where
$M^2$ transforms under SU$(2)_F$ as two-index tensor,
\beq
   M^2=\begin{pmatrix}
      \phm m_{11}^2 &\,\,\,-m_{12}^2\\
      -\overline m_{12}^2 &\,\,\, \phm m_{22}^2\end{pmatrix}\,,
\eeq
with $(M^2)_1{}^2\equiv -m_{12}^2$ and $\overline m_{12}^2\equiv (m_{12}^2)^*$.
For the gauge-invariant quadrilinear terms, we start with
$\Phi_{i\alpha}\Phi_{j\beta}\overline\Phi^{k\gamma}\overline\Phi^{\ell\delta}$,
and note that there are two ways to contract all the indices in a gauge
invariant manner.
The first invariant is,
\beq
   \Phi_{i\alpha}\Phi_{j\beta}\overline\Phi^{k\gamma}\overline\Phi^{\ell\delta}
   \epsilon^{\alpha\beta}\epsilon_{\gamma\delta}\ A^{ij}_{k\ell}=\Phi_{i\alpha}\Phi_{j\beta}\overline\Phi^{k\gamma}\overline\Phi^{\ell\delta}
\lt(\delta^\alpha_\gamma\delta^\beta_\delta-\delta^\alpha_\delta\delta^\beta_\gamma\rt)\epsilon^{ij}\epsilon_{k\ell}\, A\,,
\eeq
where
\beq 
A^{ij}_{k\ell}=-A^{ji}_{k\ell}=-A^{ij}_{\ell k}=A^{ji}_{\ell k}\,,
\eeq
is antisymmetric with respect to the separate interchange of upper and lower indices.  The antisymmetry property
of $A^{ij}_{k\ell}$
implies that only one independent element exists, $A^{ij}_{k\ell}=\epsilon^{ij}\epsilon_{k\ell}\, A$, where
\beq
 A=\tfrac{1}{8}\lt(\lambda_3-\lambda_4\rt)\,,
\eeq
which is a scalar with respect to SU$(2)_F$ transformations.

The second invariant is,
\beq
   \Phi_{i\alpha}\Phi_{j\beta}\overline\Phi^{k\gamma}\overline\Phi^{\ell\delta}
   \lt(\delta^\alpha_\gamma\delta^\beta_\delta+\delta^\alpha_\delta\delta^\beta_\gamma\rt)\Sigma^{ij}_{k\ell}
   \label{eq:Sigma}
\eeq
where
\beq
\Sigma^{ij}_{k\ell}=\Sigma^{ji}_{k\ell}=\Sigma^{ij}_{\ell k}=\Sigma^{ji}_{\ell k}\,,
\eeq
is symmetric with respect to the separate interchange of upper and lower indices.
In addition, hermiticity implies that $\Sigma^{ij}_{k\ell}=\overline{\Sigma^{k\ell}_{ij}}$.
In terms of the parameters in \eq{eq:lag}, we have
\beqa
   \Sigma^{11}_{11}&=&\tfrac{1}{4}\lambda_1,\quad
   \Sigma^{22}_{22}=\tfrac{1}{4}\lambda_2,\quad
   \Sigma^{12}_{12}=\tfrac{1}{8}(\lambda_3+\lambda_4),\nonumber\\[6pt]
   \Sigma^{22}_{11}&=&\overline{\Sigma^{11}_{22}}
   =\tfrac{1}{4}\lambda_5,\quad
   \Sigma^{12}_{11}=\overline{\Sigma^{11}_{12}}=\tfrac{1}{4}\lambda_6,\quad
   \Sigma^{22}_{12}=\overline{\Sigma^{12}_{22}}=\tfrac{1}{4}\lambda_7.
\eeqa

Since all the spurions transform as integer spin, they can be expressed as
SO(3) tensors~\cite{Ivanov:2005hg}, labeled by adjoint  SU$(2)_F$ indices $a,b,\ldots$.
The squared-mass term decomposes as $2\otimes 2=1\oplus 3$.  Explicitly,
\beq \label{M2}
M^2=2m_a^2 T_a+\mu^2\mathds{1}\,,
\eeq
where $T_a\equiv\half\sigma_a$ are the SU(2) generators, with normalization $\Tr(T_a T_b)=\half\delta_{ab}$,
and $\mathds{1}$ is the $2\times 2$ identity matrix.
In particular,
the antisymmetric part of the tensor product is the singlet that is given by the trace,
\beq \label{mudef}
   \mu^2\equiv \half\Tr(M^2)=\half\left(m^2_{11}+m^2_{22}\right).
\eeq
The symmetric part of the tensor product, denoted by $(2\otimes 2)_{\rm sym}$, is the triplet given by
\beq \label{ma}
   m^2_a=\Tr\lt(M^2 T_a\rt)=\lt(-\Re\,m^2_{12}\ ,\ \Im\,m^2_{12}\ ,\
   \half(m^2_{11}-m^2_{22})\rt).
\eeq

The quadrilinear terms transform as
$(2\otimes 2)_{\rm sym}\otimes(2\otimes 2)_{\rm sym}=1\oplus 3\oplus 5$.  Explicitly,
\beq \label{sig}
\Sigma^{ij}_{k\ell}=\half D_{ab}(T_a)\ls{k}{}^i(T_b)\ls\ell{}^j+\tfrac{1}{8}P_a\left[(T_a)\ls{k}{}^i\delta_\ell^j+(T_a)\ls{k}{}^j\delta_\ell^i
+(T_a)\ls\ell{}^j\delta_k^i+(T_a)\ls\ell{}^i\delta_k^j\right]+\tfrac{1}{24}S(\delta_k^i\delta_\ell^j+\delta_\ell^i\delta_k^j)\,,
\eeq
where $D_{ab}$ is a traceless symmetric second-rank tensor.   Using the Fierz identity,
\beqa
(T_a)\ls{k}{}^i (T_b)\ls\ell{}^j&=&\half\left[(T_a)\ls{k}{}^j (T_b)\ls\ell{}^i+(T_b)\ls{k}{}^j (T_a)\ls\ell{}^i
-\delta_{ab}(T_c)\ls{k}{}^j (T_c)\ls\ell{}^i+\tfrac{1}{4}\delta_{ab}\delta_k^j\delta_\ell^i\right]
\nonumber \\[6pt]
&&
+\tfrac{1}{4}i\epsilon_{abc}\left[\delta_\ell^i (T_c)\ls{k}{}^j-\delta_k^j(T_c)\ls\ell{}^i\right]\,,
\eeqa
it follows that $D_{ab} (T_a)\ls{k}{}^i(T_b)\ls\ell{}^j=D_{ab} (T_a)\ls{k}{}^j(T_b)\ls\ell{}^i$.  Hence, $\Sigma^{ij}_{k\ell}$ given by \eq{sig} is
symmetric under the separate interchange of its lower and its upper indices, as required.

Using \eq{sig},
the singlet is given by the trace,
\beq
   S=4\Sigma^{ij}_{ij}=
   \lambda_1+\lambda_2+\lambda_3+\lambda_4,
\eeq
the triplet is given by
\beq \label{pa}
   P_a=4\Sigma^{ij}_{k\ell}\lt(T_a\rt)\ls{j}{}^{\ell}\delta^k_i
   =\mbox{\huge (}\,\Re\lt(\lambda_6+\lambda_7\rt)\,,\,
   -\Im\lt(\lambda_6+\lambda_7\rt)\,,\,
   \half(\lambda_1-\lambda_2)\,\mbox{\huge )},
\eeq
and the 5-plet is a traceless symmetric second-rank tensor given by
\beq \label{dab}
   D_{ab} = 2\lt[4\Sigma^{ij}_{k\ell}\lt(T_a\rt)\ls{j}{}^{\ell}\lt(T_b\rt)\ls{i}{}^k
   -\tfrac{1}{3}\Sigma^{ij}_{ij}\delta_{ab}\rt]
   =\begin{pmatrix}
     -\frac{1}{3}\Delta+\Re\,\lambda_5 &\,\,\,
     -\Im\,\lambda_5 & \,\,\,\phm\Re\,\lt(\lambda_6-\lambda_7\rt)\\
     -\Im\,\lambda_5 &\,\,\,
     -\frac{1}{3}\Delta-\Re\,\lambda_5 &\,\,\, -\Im\,\lt(\lambda_6\!-\!\lambda_7\rt)\\
    \phm \Re\,\lt(\lambda_6\!-\!\lambda_7\rt) &\,\,\,
     -\Im\,\lt(\lambda_6\!-\!\lambda_7\rt)
     & \,\,\,\frac{2}{3}\Delta\end{pmatrix}\!\!,
\eeq
where $\Delta\equiv \half(\lambda_1+\lambda_2)-\lambda_3-\lambda_4$.

The above results are equivalent to the group-theoretical decomposition of the 2HDM scalar
potential obtained in \cite{Ivanov:2005hg}.\footnote{To obtain Ivanov's results~\cite{Ivanov:2005hg}, one simply
replaces $M^2$ and $\Sigma^{ij}_{k\ell}$ with their complex conjugates in the above expressions for
$m_a^2$, $P_a$ and $D_{ab}$.  Ivanov also introduces different overall normalizations for these
quantities, which are not critical to the applications presented in this Appendix.}

\subsection{Invariant relations}
Having obtained all the SU$(2)_F$ spurions, we can now
find invariant relations among parameters, \ie, relations
that hold in every basis, provided they hold in one basis.
At the linear level, invariant relations can be obtained either by setting  a singlet quantity
to a constant (any constants will do), or by setting the non-singlet spurions
to zero.  This procedure yields six invariant linear relations:
\ben
   \item $m^2_{11}+m^2_{22}=2\mu_0^2$,
   \item $\lambda_3-\lambda_4=8A_0$,
   \item $\lambda_1+\lambda_2+\lambda_3+\lambda_4=S_0$,
   \item $m^2_{11}-m^2_{22}=m^2_{12}=0$,
   \item $\lambda_1-\lambda_2=\lambda_6+\lambda_7=0$,
   \item $\half(\lambda_1+\lambda_2)-\lambda_3-\lambda_4,
   =\lambda_5=\lambda_6-\lambda_7=0$,
\een
where $\mu_0^2$, $A_0$, and $S_0$ are arbitrary constants.
This generalizes~\cite{GH}, in which relations 4 and~6 were noted and discussed.
If the scalar potential is SU$(2)_F$-invariant, then the relations 4, 5 and 6 above must
be simultaneously satisfied, \ie,
\beq
   m^2_{11}=m^2_{22}\,,\quad m^2_{12}=0\,,\quad
   \lambda_1=\lambda_2=\lambda_3+\lambda_4\,,
   \quad \lambda_5=\lambda_6=\lambda_7=0.
\eeq
This particular model was introduced previously in~\cite{GCP} and exhibits the largest allowed Higgs family symmetry of
the 2HDM scalar potential.

Higher order invariant relations may be constructed by forming scalar combinations of products
of the nontrivial spurions $m^2_a$, $P_a$, and $D_{ab}$.
For example, we can obtain invariant quadratic relations by
constructing scalar quantities from the product of two spurions and setting the result to a constant.
For example,
\beqa
   m^2_am^2_a &=& \lt|m^2_{12}\rt|^2+\tfrac{1}{4}\lt(m^2_{11}-m^2_{22}\rt)^2
   ={\rm const.}\\[6pt]
   P_aP_a &=& \lt|\lambda_6+\lambda_7\rt|^2+\tfrac{1}{4}\lt(\lambda_1-\lambda_2\rt)^2
   ={\rm const.}\\[6pt]
 m^2_aP_a&=&
   \Re \lt[m^2_{12}\lt(\lambda_6+\lambda_7\rt)\rt]
   +\tfrac{1}{4}\lt(m^2_{11}-m^2_{22}\rt)\lt(\lambda_1-\lambda_2\rt)={\rm const.}\\[6pt]
   \Tr (D^2) &=& \tfrac{1}{6}\left(\lambda_1+\lambda_2-2\lambda_2-2\lambda_4\right)^2+2|\lambda_5|^2
+2|\lambda_6-\lambda_7|^2 = {\rm const.}
\eeqa
An example of an invariant cubic relation is $m^2_aP_bD_{ab}=0$,
and so on.

\subsection{Condition for the existence of a $\boldsymbol{{\rm U}(1)_{\rm PQ}}$ symmetry}
In order to exemplify the power of SU$(2)_F$ spurion analysis, we derive
the condition for existence of a U(1) global symmetry, which in a particular basis for the
scalar fields coincides with the Peccei-Quinn symmetry and the corresponding
U(1) generator is $T^3_{ij}\id_{\alpha\beta}$ [cf.~\eq{t3}].
%
In an arbitrary basis, the generator of U(1)$_{\rm PQ}$, denoted by $T_{\rm PQ}$, must be a linear combination of the SU(2)$_F$
generators.   Hence,
\beq \label{gen}
T_{\rm PQ}=q_a T_a=\half q_a \sigma_a\,,
\eeq
which defines the three-vector $q_a$, transforming under the adjoint representation of SU(2)$_F$.
It is convenient to normalize $q_a$ such that its squared-length is $q_a q_a=1$.
If the scalar potential preserves the ${\rm U}(1)_{\rm PQ}$ symmetry,
then all spurions must be fixed (up to an overall scale) by $q_a$.  In particular,
\beq \label{pq}
m_a^2=c_1 q_a\,,\qquad P_a=c_2 q_a\,,\qquad D_{ab}=c_3\left(q_a q_b-\tfrac{1}{3}\delta_{ab}\right)\,,
\eeq
where the $c_i$ are arbitrary constants.  \Eq{pq} provides an elegant basis-independent
set of conditions for the existence of a PQ symmetry in the 2HDM.  One can use the explicit expressions
for $m_a^2$, $P_a$ and $D_{ab}$ [cf.~eqs.~(\ref{ma}), (\ref{pa}) and (\ref{dab}), respectively] to
rewrite \eq{pq} in terms of the 2HDM scalar potential parameters in an arbitrary basis.  The resulting
equations are not particularly illuminating, so we do not write them out here.

To verify the above assertion, consider the spurion, $M^2=m_a^2\sigma_a+\mu^2\mathds{1}$
introduced in \eq{M2}.
The triplet spurion $m_a^2$ transforms under the adjoint representation of SU(2)$_F$, and therefore
breaks the global SU(2)$_F$ symmetry down to U(1)$_{\rm PQ}$~\cite{Li:1973mq}.  The condition that the U(1)$_{\rm PQ}$ is
preserved is equivalent to the requirement that
\beq
[T_{\rm PQ}\,,\,M^2]=0\,.
\eeq
Inserting \eqs{M2}{gen} into the above condition yields
\beq \label{condition}
\half q_a m^2_b [\sigma_a\,,\,\sigma_b]=i\epsilon_{abc} q_a m_b^2 \sigma_c=0\,.
\eeq
\Eq{condition} implies that $q_a\propto m_a^2$, which identifies the U(1)$_{\rm PQ}$ generator.
Indeed, all triplet spurions must be proportional to $q_a$ as indicated in \eq{pq}, since any two
non-parallel triplet spurions would completely break the SU(2)$_F$ global symmetry~\cite{Li:1973mq}.
Likewise, the condition that U(1)$_{\rm PQ}$ is
conserved by the spurion $\Sigma^{ij}_{k\ell}$ is
equivalent to the requirement that
\beq
\Sigma^{mj}_{k\ell}(T_{\rm PQ})\ls{m}{}^i-\Sigma^{ij}_{n\ell}(T_{\rm PQ})\ls{k}{}^n+\Sigma^{im}_{k\ell}(T_{\rm PQ})\ls{m}{}^j-\Sigma^{ij}_{kn}(T_{\rm PQ})\ls\ell{}^n=0\,.
\eeq
Using \eq{sig}, it follows that
\beq
q_c D_{ab}\Bigl\{(T_b)\ls{k}{}^i[T_a\,,\,T_c]\ls\ell{}^j+(T_b)\ls{\ell}{}^j[T_a\,,\,T_c]\ls{k}{}^i\Bigr\}
=iq_c D_{ab}\epsilon_{ace}\left[(T_b)\ls{k}{}^i(T_e)\ls\ell{}^j+(T_b)\ls{\ell}{}^j(T_e)\ls{k}{}^i
\right]=0\,,
\eeq
which is satisfied by $D_{ab}\propto q_a q_b-\tfrac{1}{3} \delta_{ab}$ as indicated in \eq{pq}.
\clearpage

One is always free to choose a convenient basis for the scalar fields of the 2HDM
by diagonalizing $D_{ab}$.  The eigenvalues of $D_{ab}=c_3(q_a q_b-\tfrac{1}{3} \delta_{ab})$ are
$
-\tfrac{1}{3}c_3 \,,\, -\tfrac{1}{3}c_3 \,,\,
+\tfrac{2}{3}c_3
$
(note the doubly-degenerate eigenvalue assuming that $c_3\neq 0$).  It is straightforward to check that
$D_{ab}$ is diagonal when $q_a=(0,0,1)$.  Then \eq{pq} implies that $m_{12}^2=\lambda_5=\lambda_6=\lambda_7=0$ in the $D$-diagonal basis,
which yields the standard form for the 2HDM scalar potential with PQ-symmetry $\Phi_1\to e^{i\alpha} \Phi_1$
and $\Phi_2\to e^{-i\alpha}\Phi_2$.  Moreover, in the $D$-diagonal basis, we can identify $c_1=\half(m_{11}^2-m_{22}^2)$,
$c_2=\half(\lambda_1-\lambda_2)$ and $c_3=\Delta=\half(\lambda_1+\lambda_2)-\lambda_3-\lambda_4$.

Of course, \eq{pq} is applicable in an arbitrary basis.  These conditions are equivalent to the
invariant conditions given in \cite{DH,GCP}, although the formulation of \eq{pq} is much simpler and
transparent than the conditions originally given.  Note that at the exceptional point of parameter space
identified in \cite{DH} where $m_{11}^2=m_{22}^2$, $m_{12}^2=0$, $\lambda_1=\lambda_2$ and
$\lambda_7=-\lambda_6$, it follows that $m_a^2=P_a=0$.  In this case, the condition for PQ symmetry
is simply the existence of a doubly-degenerate eigenvalue of $D_{ab}$ as first noted in \cite{GCP}.
This latter condition implies that $D_{ab}\propto q_a q_b-\tfrac{1}{3}\delta_{ab}$ for some unit
vector $q_a$, which then determines the PQ generator given in \eq{gen}.

\end{document}